\documentstyle[preprint]{aastex}

\def\spose#1{\hbox to 0pt{#1\hss}}
\newcommand\lsim{\mathrel{\spose{\lower 3.0pt\hbox{$\mathchar"218$}}
     \raise 2.0pt\hbox{$\mathchar"13C$}}}
\newcommand\gsim{\mathrel{\spose{\lower 3.0pt\hbox{$\mathchar"218$}}
     \raise 2.0pt\hbox{$\mathchar"13E$}}}
\newcommand\half{\frac{1}{2}}

\shortauthors{J. SHIN \& S. S. KIM}
\shorttitle{ADI METHOD FOR 2D FOKKER-PLANCK EQUATION}

\begin{document}
\title{ALTERNATING DIRECTION IMPLICIT METHOD FOR TWO-DIMENSIONAL FOKKER-PLANCK
EQUATION OF DENSE SPHERICAL STELLAR SYSTEMS}
\author{Jihye Shin and Sungsoo S. Kim\altaffilmark{1}}
\affil{Kyung Hee University, Dept. of Astronomy \& Space Science,
Yongin, Kyungki 446-701, Korea}
\altaffiltext{1}{Corresponding author; sungsoo.kim@khu.ac.kr}

\begin{abstract}
The Fokker-Planck (FP) model is one of the commonly used methods for studies
of the dynamical evolution of dense spherical stellar systems such as globular
clusters and galactic nuclei.  The FP model is numerically stable in most
cases, but we find that it encounters numerical difficulties rather often
when the effects of tidal shocks are included in two-dimensional (energy and
angular momentum space) version of the FP model or when the initial
condition is extreme (e.g., a very large cluster mass and a small cluster
radius).  To avoid such a problem, we have developed a new integration
scheme for a two-dimensional FP equation by adopting an Alternating
Direction Implicit (ADI) method given in the Douglas-Rachford split form.
We find that our ADI method reduces the computing time by a factor of
$\sim 2$ compared to the fully implicit method, and resolves problems
of numerical instability.
\end{abstract}


\section{INTRODUCTION}

The Fokker-Planck (FP) model is a statistical way of describing the time
evolution of a probability density function under the effects of drift
and diffusion.  One of the first uses of FP model in stellar dynamics
was by Cohn (1979), who developed a numerical method that directly
integrates the two-dimensional (2D) FP equation in energy-angular
momentum ($E$, $J$) space targeted for dense spherical stellar systems
such as globular clusters.  But this pioneering attempt suffered
a non-negligible numerical problem with the energy conservation, and
to eliminate this problem, Cohn (1980) developed a one-dimensional FP
model in energy space with an assumption that the velocity distribution
of stars in the cluster is isotropic (i.e., the distribution function
can be described by energy only).  He adopted the finite-difference
scheme by Chang and Cooper (1970) and was able to greatly reduce the
numerical errors.

As the computing power of workstations greatly increased in 1990's,
it became feasible to integrate the 2D-FP equation with a relatively
large number of grids on workstations.  Takahashi (1995) challenged
the 2D-FP model again with the help of the increased computing power,
and successfully developed a numerically reliable method for 2D-FP
equation by adopting the Chang \& Cooper finite-difference scheme.
Takahashi found that this scheme greatly reduced the numerical errors
when applied to the energy dimension, but it was not
so effective when applied to the angular momentum space.  Thus he applied
the scheme only to the energy space.  Although his model was not a full 2D
generalization of the Chang \& Cooper scheme, it significantly reduced
the numerical errors on the energy conservation compared to the 2D model
by Cohn (1979).

The 2D-FP model developed by Takahashi (1995) reliably calculates the
dynamical evolution of dense stellar systems in most cases, but we find
that it encounters a numerical problem when the effect of tidal shocks
(disk and/or bulge shocks) is added to the model or when the initial
condition is extreme (e.g., a very large cluster mass and a small
cluster radius).  The Chang \& Cooper
scheme adopted by Takahashi is an implicit finite-difference method,
which involves solving a large matrix.  A matrix inversion is a
numerically challenging task particularly when the magnitude range
of the numbers in the matrix is large, and in such a case, the inversion
can result in significantly inaccurate answers.  We find that when
the effect of tidal shocks is added, Takahashi's model encounters a
numerical problem in the matrix inversion part of the implicit scheme
and results in a partially negative distribution function.

In the present paper, we develop a new numerical method for the 2D-FP
equation by adopting an Alternating Direction Implicit (ADI) method,
instead of the fully implicit method as in the Chang \& Cooper's scheme,
to overcome the forementioned numerical problem.  Fokker-Planck models
are advantageous over N-body simulations particularly when studying the
dynamical evolution of a system of globular clusters or galactic nuclei,
and the method presented here will be useful as a numerically efficient
and stable tool for such studies.

We briefly introduce the formulation of the 2D-FP equation and show
its finite-difference expressions in implicit and explicit fashions
in Section 2.  We present and discuss the ADI formulation of the 2D-FP
equation in Section 3, and summarize our findings in Section 4.

\section{TWO-DIMENSIONAL FOKKER-PLANCK EQUATION}

Here we briefly introduce the formulation of the 2D-FP equation following
the discription by Takahashi (1995).
In a steady-state spherical system, a distribution
function $f(\vec r, \vec v, t)$ with velocity space $\vec v$, volume space
$\vec r$ at time $t$ is a function of only energy $E$ and angular
momentum $J$ per unit mass, and it evolves only due to collisional
effects.  The evolution of $f$ can be
described by the orbit-averaged FP equation in ($E$,$J$)-space, because the
relaxation (or diffusion) time scale is much longer than the dynamical
time scale.  The scaled angular momentum $R$ is often used instead of $J$ as
a basic variable, and is defined as $J^2/J^2_c(E)$ where $J_c(E)$ is
the angular momentum of a circular orbit of energy $E$.  The number density
$N(E,R)$ in ($E$, $R$)-space is
given by
\begin{eqnarray}
	N(E,R) & =      & 4\pi^2 P(E,R) J^2_c(E) f(E,R),\\ \nonumber
	       & \equiv & A(E,R) f(E,R),
\end{eqnarray}
where $P(E,R)$ is the orbital period.
When the gravitational potential is fixed, the 2D-FP equation can be
written in a flux-conserving form (Cohn 1979) such that
\begin{equation}
        A \frac{\partial f}{\partial t} = - \frac{\partial F_E}{\partial E}
                                          - \frac{\partial F_R}{\partial R},
\end{equation}
where
\begin{eqnarray}
        -F_E & = &   D_{EE} \frac{\partial f}{\partial E}
                   + D_{ER} \frac{\partial f}{\partial R}
                   + D_E f \nonumber \\
        -F_R & = &   D_{RE} \frac{\partial f}{\partial E}
                   + D_{RR} \frac{\partial f}{\partial R}
                   + D_R f,
\end{eqnarray}
and the expressions for the diffusion coefficients $D$'s are given in Appendix
C of Cohn (1979).

The implicit version of the finite-difference formulation for the above
2D-FP equation can be written as
\begin{equation}
\label{implicit1}
  A^n_{i,j}\frac{f^{n+1}_{i,j} - f^n_{i,j}}{\Delta t}
    = -\frac{\tilde F_{x\, i+\half,j} - \tilde F_{x\, i-\half,j}}{\Delta x}
      -\frac{\tilde F_{y\, i,j+\half} - \tilde F_{y\, i,j-\half}}{\Delta y},
\end{equation}
where
\begin{eqnarray}
  \tilde F_{x\, i+\half,j}
    & = & -D^n_{x\, i+\half,j} \tilde f_{i+\half,j}
          -D^n_{xx\, i+\half,j}
            \frac{\tilde f_{i+1,j} - \tilde f_{i,j}}{\Delta x}
          -D^n_{xy\, i+\half,j}
            \frac{\tilde f_{i+\half,j+\half} - \tilde f_{i+\half,j-\half}}
                 {\Delta y} \nonumber \\
  \tilde F_{x\, i-\half,j}
    & = & -D^n_{x\, i-\half,j} \tilde f_{i-\half,j}
          -D^n_{xx\, i-\half,j}
            \frac{\tilde f_{i,j} - \tilde f_{i-1,j}}{\Delta x}
          -D^n_{xy\, i-\half,j}
            \frac{\tilde f_{i-\half,j+\half} - \tilde f_{i-\half,j-\half}}
                 {\Delta y} \nonumber \\
  \tilde F_{y\, i,j+\half}
    & = & -D^n_{y\, i,j+\half} \tilde f_{i,j+\half}
          -D^n_{yy\, i,j+\half}
            \frac{\tilde f_{i,j+1} - \tilde f_{i,j}}{\Delta y}
          -D^n_{yx\, i,j+\half}
            \frac{\tilde f_{i+\half,j+\half} - \tilde f_{i-\half,j+\half}}
                 {\Delta x} \nonumber \\
  \tilde F_{y\, i,j-\half}
    & = & -D^n_{y\, i,j-\half} \tilde f_{i,j-\half}
          -D^n_{yy\, i,j-\half}
            \frac{\tilde f_{i,j} - \tilde f_{i,j-1}}{\Delta y}
          -D^n_{yx\, i,j-\half}
            \frac{\tilde f_{i+\half,j-\half} - \tilde f_{i-\half,j-\half}}
                 {\Delta x}.
\end{eqnarray}
Here, $f^n_{i,j}$ is the distribution function at the energy and angular
momentum mesh of index ($i$,$j$) in the $n$-th time step, and $\Delta x$,
$\Delta y$, and $\Delta t$ are the intervals of energy mesh, angular momentum
mesh, and time, respectively.  Takahashi (1995) adopts the Crank-Nicolson
scheme for the time advance and the cross terms, i.e.
\begin{eqnarray}
	\tilde f_{i,j} & = & \half (f^n_{i,j}+f^{n+1}_{i,j}) \nonumber \\
	f_{i\pm\half ,j\pm\half} & = & \frac{1}{4} (f_{i,j}+f_{i\pm1,j}+
                                       f_{i,j\pm1}+f_{i\pm1,j\pm1}),
\end{eqnarray}
while he adopts the Chang \& Cooper scheme for the energy dimension such that
\begin{eqnarray}
	f_{i+\half ,j} & = & \delta_{x\, i,j} f_{i,j} +
                             (1-\delta_{x\, i,j}) f_{i+1,j}   \nonumber \\
	\delta_{x\, i,j}&= & \frac{1}{w_{x\, i,j}}
                             - \frac{1}{\exp (w_{x\, i,j})-1} \nonumber \\
	w_{x\, i,j} & = & \Delta x \frac{D_{x\, i+\half,j}}{D_{xx\, i+\half,j}}.
\end{eqnarray}
For the angular momentum dimension, $\delta_y$ is always set to be 0.5.

Rearranging equation (\ref{implicit1}) for $f^{n+1}_{i,j}$ results in
\begin{equation}
\label{implicit2}
	\Sigma^{+1}_{l=-1}\Sigma^{+1}_{m=-1} B_{i+l,j+m}f^{n+1}_{i+l,j+m}
	+ \frac{A^n_{i,j}}{\Delta t} B_{i,j}f^{n+1}_{i,j}
		=
	- \Sigma^{+1}_{l=-1}\Sigma^{+1}_{m=-1} B_{i+l,j+m}f^n_{i+l,j+m}
	+ \frac{A^n_{i,j}}{\Delta t} B_{i,j}f^n_{i,j},
\end{equation}
where
\begin{eqnarray}
  B_{i-1,j-1} & = & -\frac{ D^n_{xy\, i-\half,j}
                           +D^n_{yx\, i,j-\half}}{4\Delta x\Delta y} \nonumber\\
  B_{i-1,j} & = &  \frac{\delta_{x\, i-1,j} D^n_{x\, i-\half,j}}{\Delta x}
                    -\frac{D^n_{xx\, i-\half,j}}{\Delta x\Delta x}
                    -\frac{ D^n_{yx\, i,j-\half}
                           -D^n_{yx\, i,j+\half}}{4\Delta x\Delta y} \nonumber\\
  B_{i-1,j+1} & = &  \frac{ D^n_{xy\, i-\half,j}
                           +D^n_{yx\, i,j+\half}}{4\Delta x\Delta y} \nonumber\\
  B_{i  ,j-1} & = &  \frac{\half D^n_{y\, i,j-\half}}{\Delta y}
                    -\frac{D^n_{yy\, i,j-\half}}{\Delta y\Delta y}
                    -\frac{ D^n_{xy\, i-\half,j}
                           -D^n_{xy\, i+\half,j}}{4\Delta x\Delta y} \nonumber\\
  B_{i  ,j} & = &  \frac{ (1-\delta_{x\, i-1,j}) D^n_{x\, i-\half,j}
                           -\delta_{x\, i,j} D^n_{x\, i+\half,j}}{\Delta x}
                    +\frac{ \half D^n_{y\, i,j-\half}
                           -\half D^n_{y\, i,j+\half}}{\Delta y}   \nonumber\\
              &   & +\frac{D^n_{xx\, i-\half,j}+
                           D^n_{xx\, i+\half,j}}{\Delta x\Delta x}
                    +\frac{D^n_{yy\, i,j-\half}+
                           D^n_{yy\, i,j+\half}}{\Delta y\Delta y} \nonumber\\
  B_{i,  j+1} & = & -\frac{\half D^n_{y\, i,j+\half}}{\Delta y}
                    -\frac{D^n_{yy\, i,j+\half}}{\Delta y\Delta y}
                    +\frac{ D^n_{xy\, i-\half,j}
                           -D^n_{xy\, i+\half,j}}{4\Delta x\Delta y} \nonumber\\
  B_{i+1,j-1} & = &  \frac{ D^n_{xy\, i+\half,j}
                           +D^n_{yx\, i,j-\half}}{4\Delta x\Delta y} \nonumber\\
  B_{i+1,j} & = & -\frac{(1-\delta_{x\, i,j}) D^n_{x\, i+\half,j}}{\Delta x}
                    -\frac{D^n_{xx\, i+\half,j}}{\Delta x\Delta x}
                    +\frac{ D^n_{yx\, i,j-\half}
                           -D^n_{yx\, i,j+\half}}{4\Delta x\Delta y} \nonumber\\
  B_{i+1,j+1} & = & -\frac{ D^n_{xy\, i+\half,j}
                           +D^n_{yx\, i,j+\half}}{4\Delta x\Delta y}.
\end{eqnarray}
Equation (\ref{implicit2}) forms a set of linear equations and its solution
can be obtained by inverting the matrix whose components are $B_{i,j}$.
Because every component of $B_{i,j}$ is non-zero in general, and because
the size of the matrix easily goes over 50 in each dimension, solving
equation (\ref{implicit2}) becomes a numerically challenging task.

On the other hand, the explicit version of the finite-difference formulation
for the 2D-FP equation can be written as
\begin{equation}
	A_{i,j}^{n} \frac{f_{i,j}^{n+1}-f_{i,j}^{n}}{\Delta t}
		= - \frac{F^n_{x\,i+\half,j}-F^n_{x\,i-\half,j}}{\Delta x}
		  - \frac{F^n_{y\,i,j+\half}-F^n_{x\,i,j-\half}}{\Delta y},
\end{equation}
where
\begin{equation}
	F^n_{x\,i+\half,j}
		= - D_{x\,i+\half,j}^n f^n_{i+\half,j}
		  - D_{xx\,i+\half,j}^n \frac{f^n_{i+1,j}-f^n_{i,j}}{\Delta x}
		  - D_{xy\,i+\half,j}^n \frac{f^n_{i+\half,j+\half}
		    -f^n_{i+\half,j-\half}}{\Delta y},
\end{equation}
and $F^n_{x\,i-\half,j}$, $F^n_{x\,i,j+\half}$, and $F^n_{x\,i,j-\half}$ are
similarly defined.  Since the $f^{n+1}_{i,j}$ term appears only once in
the above formulation, an inversion of a matrix is not involved in obtaining
the solution at the next step.

\section{ALTERNATING DIRECTION IMPLICIT METHOD}

As shown in Section 2, implicit finite-difference methods obtain the
solution for the next
time step from the state of both current and next time steps, while
explicit methods obtain the solution from the state of the current
time step only.  Implicit methods require more computations per step
but they can implement longer time intervals without suffering numerical
instabilities (note that, however, implicit methods are stable for
one-dimensional problems, but not necessarily for multi-dimensional
problems).  Implicit methods
are preferred in most cases because of this benefit, but they involve
the inversion of a matrix, which can be numerically problematic in some
cases.  When such a problem is encountered, one could implement an explicit
method instead, but explicit methods requires much smaller time intervals
than an implicit method to avoid numerical instabilities.  We find that
the required small time intervals greatly increase the computing time
to the degree that the merit of the FP model over direct N-body
simulations is lost.

The ADI method is a finite-difference method for solving differential
equations in two or more dimensions.  For a 2D problem, the ADI method
solves the first dimension implicitly and the second dimension explicitly,
and in the next step the first dimension explicitly and the second
dimension implicitly, and so on.  This method is unconditionally stable,
and since it applies the implicit scheme to one dimension at a time,
the non-zero terms are present only in the three diagonal lines in the
matrix, which is considerably simpler to solve compared to the matrix
created by the fully implicit method (such as the Chang \& Cooper method)
in 2D.

In the present paper, we develop an ADI-type finite difference method for
solving a 2D-FP equation for dense spherical stellar systems.  We adopt
an ADI scheme in Douglas \& Rachford (1956) split form and write the
finite-difference formulation such that
\begin{eqnarray}
  \left [   \frac{A^n_{i,j}}{\Delta t}
          - \frac{\delta^2_x}{2 (\Delta x)^2}
          + \frac{\nabla_x}{2 \Delta x}
  \right ] f^{n+1^*}_{i,j}
  & = &
  \left [ \frac{A^n_{i,j}}{\Delta t}
          + \frac{\delta^2_y}{(\Delta y)^2}
          - \frac{\nabla_y}{\Delta y}
          + \frac{\delta^2_x}{2 (\Delta x)^2}
          - \frac{\nabla_x}{2 \Delta x}
          + \frac{\delta_{xy}}{\Delta x \Delta y}
  \right ] f^n_{i,j} \nonumber \\
  \left [   \frac{A^n_{i,j}}{\Delta t}
          - \frac{\delta^2_y}{2 (\Delta y)^2}
          + \frac{\nabla_y}{2 \Delta y}
  \right ] f^{n+1}_{i,j}
  & = &
  \frac{A^n_{i,j}}{\Delta t} f^{n+1^*}_{i,j}
  - \left [   \frac{\delta^2_y}{2 (\Delta y)^2}
            - \frac{\nabla_y}{2 \Delta y}
  \right ] f^n_{i,j},
\end{eqnarray}
where
\begin{eqnarray}
  \delta^2_x f^n_{i,j} & = &
        D^n_{xx\, i+\half,j} (f^n_{i+1,j}-f^n_{i,j})
      - D^n_{xx\, i-\half,j} (f^n_{i,j}-f^n_{i-1,j}) \nonumber \\
  \delta^2_y f^n_{i,j} & = &
        D^n_{yy\, i,j+\half} (f^n_{i,j+1}-f^n_{i,j})
      - D^n_{yy\, i,j-\half} (f^n_{i,j}-f^n_{i,j-1}) \nonumber \\
  \delta_{xy} f^n_{i,j} & = &
        \frac{1}{4} [
        D^n_{xy\, i+\half,j} (f^n_{i+\half,j+\half} - f^n_{i+\half,j-\half})
      - D^n_{xy\, i-\half,j} (f^n_{i-\half,j+\half} - f^n_{i-\half,j-\half})
        \nonumber \nonumber \\
  & & + D^n_{yx\, i,j+\half} (f^n_{i+\half,j+\half} - f^n_{i-\half,j+\half})
      - D^n_{yx\, i,j-\half} (f^n_{i+\half,j-\half} - f^n_{i-\half,j-\half})
        ] \nonumber \\
  \nabla_x f^n_{i,j} & = &
      \half [   D^n_{x\, i+\half,j} (f^n_{i+1,j}+f^n_{i,j})
              - D^n_{x\, i-\half,j} (f^n_{i,j}+f^n_{i-1,j})
            ] \nonumber \\
  \nabla_y f^n_{i,j} & = &
      \half [   D^n_{y\, i,j+\half} (f^n_{i,j+1}+f^n_{i,j})
              - D^n_{y\, i,j-\half} (f^n_{i,j}+f^n_{i,j-1})
            ].
\end{eqnarray}
Here, $f^n_{i+\half,j+\half}$ and similar expressions are the distribution
functions at the center of the four nearby mesh points.  For example,
\begin{equation}
	f^n_{i+\half,j+\half} = \frac{1}{4} ( f^n_{i,j} + f^n_{i+1,j} +
	                                      f^n_{i,j+1} + f^n_{i+1,j+1} ).
\end{equation}
For the boundary conditions, we impose $D=0$ at the boundary meshes.
An example of the boundary conditions at $i=1$
(the first mesh point in the energy dimension) is
\begin{eqnarray}
  \delta^2_x f^n_{1,j} & = & D^n_{xx\, 1+\half,j} (f^n_{2,j}-f^n_{1,j})
                                                                   \nonumber \\
  \delta^2_y f^n_{1,j} & = & 
        D^n_{yy\, 1,j+\half} (f^n_{1,j+1}-f^n_{1,j})
      - D^n_{yy\, 1,j-\half} (f^n_{1,j}-f^n_{1,j-1}) \nonumber \\
  \delta_{xy} f^n_{1,j} & = &
        \frac{1}{4} 
        D^n_{xy\, 1+\half,j} (f^n_{1+\half,j+\half} - f^n_{1+\half,j-\half})
        \nonumber \\
  \nabla_x f^n_{1,j} & = &
        \frac{1}{2} D^n_{x\, 1+\half,j} (f^n_{2,j}+f^n_{1,j}) \nonumber \\
  \nabla_y f^n_{1,j} & = & 
        \half [   D^n_{y\, 1,j+\half} (f^n_{1,j+1}+f^n_{1,j})
                - D^n_{y\, 1,j-\half} (f^n_{1,j}+f^n_{1,j-1})
              ].
\end{eqnarray}
The implicit scheme is first applied to the E-direction to obtain
$f^{{n+1}^*}_{i,j}$, then applied to the R-direction to obtain
the solution at the next step,
$f^{n+1}_{i,j}$, with the information of $f^{{n+1}^*}_{i,j}$.  Solving for
$f^{{n+1}^*}_{i,j}$ and $f^{n+1}_{i,j}$ each requires an inversion of a
tridiagonal matrix, which is a numerically straightforward task with
only minimal numerical errors.  We find that our ADI method requires
$\sim 50$~\% less computing time than the full implicit method by Takahashi
(1995) when mesh points of 181, 51, and 151 are used for energy, angular
momentum, and radial meshes, respectively.

More importantly, our ADI method perfectly prevents numerical problems
encountered by the fully implicit method.  We performed 2D-FP calculations
for 578 different initial conditions (different cluster masses, galactocentric
radii, orbit eccentricities, and orbit inclnations relative to the galactic
plane) of globular clusters with the effects of stellar evolution, binary
heating, disk/bulge shocks, realistic orbital motions, and dynamical
friction using both implicit and ADI methods (the results of these
calculations are to be reported elsewhere).  We adopted the 2D-FP model
by Takahashi et al. (1997) and modified it for tidal binary heating,
realistic cluster orbit, dynamical friction, and disk/bulge shocks.
For disk/bulge shocks, we adopted the recipes for the heating in energy
dimension by Gnedin et al. (1999a,b) and extended them for the
energy-angular momentum space (this extention will be reported elsewhere).
The original 2D-FP model by Takahashi et al. implements an implicit method
(Chang \& Cooper scheme) for integrating the FP equation, and we modified
their model so that it can implement our ADI method instead of the
implicit method as an option.
We find that $\sim 70$~\% of the calculations performed with the implicit
method (Chang \& Cooper scheme) encountered numerical problems (negative
distribution functions or crashes during the matrix inversion) when
the effects of disk/bulge shocks are included in the calculation.
The disk/disk shocks heat the stars near the tidal boundary the most
(see Fig. 1), and it appears that inverting the matrix created by the
implicit formulism
becomes numerically difficult when the stars near the tidal boundary
are heated significantly enough.  When the effects of disk/bulge shocks
are not included, less than 10~\% of the calculations encountered
numerical problems, and these happen mostly for clusters with a very large
initial mass and/or a small initial radius.
On the other hand, none of the calculations performed with our ADI method
encountered such problems.  This clearly shows that our ADI method not only
reduces the computing time but also resolves numerical problems involved
in the fully implicit finite-difference method for the 2D-FP equation of
dense spherical stellar systems.

As an example, Fig. 2 compares the distribution functions calculated with
the ADI and implicit methods at the epoch when the implicit method encounters
a numerical problem in one of the 578 calculations discussed above.
The distribution function for the next time step
obtained with the implicit method has mostly negative values and more
importantly, it is significantly different from that of the current step.
This indicates that the matrix created by the implicit method is numerically
challenging and the matrix inversion results in a considerably incorrect
answer.  On the other hand, the solution obtained with our ADI method is
very close to the value at the current time step and does not have negative
values, implying that the ADI method is numerically stable.

\section{SUMMARY}

We have developed a new integration method for the 2D-FP equation of dense
spherical stellar systems by adopting an ADI finite-difference scheme.
This method shortens the computing time by a factor of 2 compared to
the implicit method, and does not encounter numerical problems
such as negative distribution functions or crashes during the matrix
inversion that implicit methods suffer when the effects of disk/bulge
shocks are included in the calculation or when extreme initial conditions
such as very high cluster masses and/or small cluster radii are used.
Disk/bulge shocks heat the stars near the tidal boundary of the cluster
the most, and it appears that inverting the matrix created by the implicit
formulism becomes numerically difficult when the stars near the tidal
boundary are heated significantly enough.  The ADI method applies the implicit
scheme to one dimension of the distribution function at a time and it only
needs to solve two tridiagonal matrices each time step, which is a
numerically straightforward task.  We find that this merit of the ADI
method effectively removes the problems involved with the implicit methods
such as the Chang \& Cooper scheme.

\acknowledgments
We thank Hyung Mok Lee and Koji Takahashi for helpful discussion.
This work was supported by Korea Research Foundation Grant funded by
Korea Government (MOEHRD, Basic Reasearch Promotion Fund;
KRF-2005-015-C00186), and by the Astrophysical Research Center for
the Structure and Evolution of the Cosmos (ARCSEC) of the Korea Science and
Engineering Foundation through the Science Research Center (SRC) program.
This work was in part supported by the BK21 program as well.


\clearpage
\begin{figure}
\epsscale{0.8}
\plotone{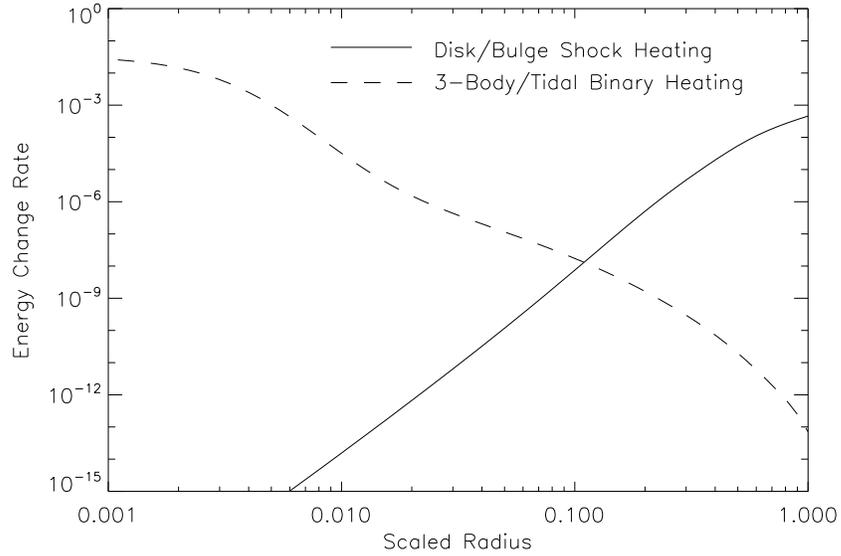}
\caption{Phase-averaged first-order energy change rates
$\langle \Delta E \rangle$ by the
disk and bulge shocks (solid) and by the three-body and tidal binary heatings
(dashed) at the epoch when the implicit method encounters a numerical
problem ($t=8.86 t_{relax}$; $t_{relax}$ is the initial half-mass relaxation
time) in one of our 578 calculations for globular clusters with the effects
of disk/bulge shocks.  Energy change rates are in arbitrary units and the
radius is in units of the tidal radius.  While the binaries preferentially
heat the core of the cluster because of the high density there, the shocks
preferentially heat the outskirt of the cluster because the tidal force by
the external gravitational field is proportional to the distance from the
cluster center.}
\end{figure}

\clearpage
\begin{figure}
\epsscale{0.8}
\plotone{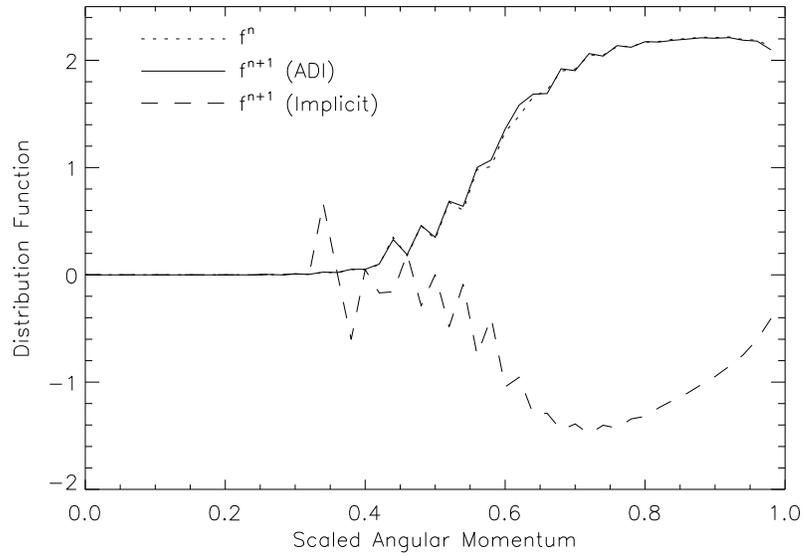}
\caption{Distribution functions of the current step (dotted) and
the next step by our ADI method (solid) and by an implicit method (dashed)
at $i=150$ for the calculation shown in Fig. 1.  The mesh point
$i=150$ is the 31st-smallest energy mesh among a total of 181 energy meshes.
The distribution functions are in arbitrary units.}
\end{figure}

\end{document}